\documentclass[twocolumn,pre,showpacs,superscriptaddress]{revtex4}

\usepackage{amsmath,amssymb}
\usepackage{color}
\usepackage{hyperref}
\usepackage{graphicx}
\usepackage{mathrsfs}
\hypersetup{colorlinks=true,linkcolor=blue}

\begin{document}

\title{Numerical approach to unbiased and driven generalized elastic model}

\author{M. Ghasemi Nezhadhaghighi}
\affiliation{Department of Physics, Sharif University of Technology, Tehran,
P.O.Box: 11365-9161, Iran}
\affiliation{Institute for Physics and Astronomy, University of Potsdam,
14476 Potsdam-Golm, Germany}
\author{A. Chechkin}
\affiliation{Max-Planck Institute for Physics of Complex Systems, Noethnitzer
Stra{\ss}e 38, 01187 Dresden, Germany}
\affiliation{Institute for Theoretical Physics, NSC KIPT, ul. Akademicheskaya
1, UA-61108 Kharkov, Ukraine}
\author{R. Metzler}
\affiliation{Institute for Physics and Astronomy, University of Potsdam,
14476 Potsdam-Golm, Germany}
\affiliation{Physics Department, Tampere University of Technology, Tampere,
Finland}

\date{\today}

\begin{abstract}
From scaling arguments and numerical simulations, we investigate the properties
of the generalized elastic model (GEM), that is used to describe various
physical systems such as polymers, membranes, single-file systems, or rough
interfaces. We compare analytical and numerical results for the subdiffusion
exponent $\beta$ characterizing the growth of the mean squared displacement
$\langle(\delta h)^2\rangle$ of the field $h$ described by the GEM dynamic
equation. We study the scaling properties of the $q$th order moments $\langle|
\delta h|^q\rangle$ with time, finding that the interface fluctuations show no
intermittent behavior. We also investigate the ergodic properties of the process
$h$ in terms of the ergodicity breaking parameter and the distribution of the
time averaged mean squared displacement. Finally, we study numerically the driven
GEM with a constant, localized perturbation and extract the characteristics of
the average drift for a tagged probe.
\end{abstract}


\maketitle

\section{Introduction}

In the past two decades, considerable theoretical and numerical effort has been
put into the characterization and quantitative modeling of stochastic patterns
such as surface growth processes \cite{surface1,surface2}, spatiodynamic profiles
of elastic chains \cite{granek1997}, single-file systems \cite{single}, membranes
\cite{zilman2002,helfer2000}, and
polymers \cite{doi,amblard1996,everaers1999,caspi1998}, as well as fluid
interface motion through porous media \cite{buldyrev1992,nissen2001}, the shape
of vortex lines in high $T_c$ superconductors \cite{bustingorry2006}, tumor
growth \cite{bru1998}, and crack propagation \cite{podlubny}. For obvious
reasons, these processes are of substantial interest both from a fundamental
physics and technological applications point of view. To obtain a quantitative
understanding, different continuum models have been proposed and studied to
reproduce the dynamics of such natural phenomena. The simplest and well-known
examples are the Edwards-Wilkinson and the Mullins-Herring equations
\cite{surface1,surface2}. Such models provide information about the
out-of-equilibrium dynamics of the field $h(\mathbf{x},t)$, that describes the
height profile of the surface under consideration, a membrane, etc. For processes
such as the spatiotemporal evolution of a polymer configuration, $h$ becomes a
vector field. In what follows, we concentrate on scalar fields $h$ and its
governing diffusion-noise equation \cite{vankampen}.

The generalized elastic model (GEM) proposed and analyzed in
Refs.~\cite{taloni2010,taloni2012,taloni2011,taloni2013} unifies various classes
of stochastic processes such as the configuration dynamics of semiflexible,
flexible, and Rouse polymers, fluid membranes, single-file system, fluctuating
interfaces, solid surfaces, and the diffusion-noise equation. Suppose you follow
the dynamics of a particular tracer particle in a stochastic system described by
the field $h(\mathbf{x},t)$. This could be a labeled lipid molecule in a membrane
or an individual particle in a single-file system. The motion of such a tracer
particle is then necessarily coupled to the rest of the system, and this
correlated dynamics effects the subdiffusive motion of the tracer particle,
characterized by the subdiffusion exponent $\beta$ in the mean squared
displacement of the field $h$ with time,
\begin{equation}
\langle(\delta h)^2\rangle\simeq t^{2\beta},
\end{equation}
with $0<\beta<1/2$ \cite{REM}
The dynamic exponent $\beta$ is but one of three scaling exponents characteristic
for stochastic processes described by the GEM, the other two being the roughness
exponent $\xi$ and the dynamic exponent $\nu$. The triple of these exponents are
most commonly used to classify surface growth dynamics \cite{surface1,surface2}.
Here we investigate numerically the scaling properties of the GEM, in particular,
we obtain the dynamic exponent $\tau(q,\beta)$ of the general $q$th order moments
$\langle|\delta h|^q\rangle\simeq t^{\tau(q,\beta)}$.

Starting from early studies of the long-time out-of-equilibrium dynamics of
glassy materials \cite{struik1978}, many complex systems characterized by
anomalous diffusion \cite{bouchaud,report}
were shown to exhibit ageing effects and weak ergodicity
breaking \cite{bouchaud1992,cugliandolo1995,monthus1996,rinn2000,barkai2003,%
bel2005,rebenshtok2007,johannes}. Respectively, these effects refer to
the dependence of the dynamics of such system on their age since the
initial preparation, and the fact that long time and ensemble averaged
observables behave differently and are irreproducible. In particular, important
consequences of such weak ergodicity breaking were studied in the non-stationary
continuous time random walk (CTRW) model and used to interpret single molecule
tracking data \cite{He2008,burov2011,Joen2011}. Similar weak ergodicity breaking
is observed for regular diffusion equation dynamics with space-dependent
diffusion coefficient and explicitly aging CTRW processes
\cite{cherstvy2013,lomholt}. Closely related to the GEM,
other anomalous diffusion systems such as fractional Brownian motion and
fractional Langevin equation motion are ergodic \cite{deng2009}, but exhibit
transient aging and weak ergodicity breaking \cite{jochen}. As originally
pointed out by Taloni \textit{et al.} \cite{taloni2012}, time and ensemble
averages of the squared displacement of a tracer particle in the GEM with
non-equilibrium initial conditions are disparate. In the present paper we study
numerically the ergodic properties of the GEM by probing quantities such as the
amplitude scatter of time averaged observables and the ergodicity breaking
parameter EB.

In order to further characterize the viscoelastic properties of the system under
study, we also consider the case of a driven GEM, that is, the response of the
GEM dynamics to an external localized force, supposed to act only on a single
tagged probe \cite{taloni2011}. Below we analyze the driven GEM numerically in
order to investigate the motion of this tagged probe.

The paper is organized as follows: in Section \ref{Chap:Def} we introduce the
notation and define the GEM and the GEM with localized perturbation. In Section
\ref{Chap:simulation} we report a general method to simulate the GEM numerically.
The numerical results are discussed in Section \ref{Chap:Results}. Finally, we
draw our Conclusions in Section \ref{Chap:Conclusions}. To be self-explanatory
we add two Appendices to explain efficient ways to approximate the space
fractional operator and to generate fractional Gaussian noise.

\section{Definitions and settings}\label{Chap:Def}

The GEM is defined in terms of the stochastic linear partial integrodifferential
equation \cite{taloni2010}
\begin{eqnarray}
\label{GEM}
\frac{\partial}{\partial t}h(\mathbf{x},t) = \int d^d x^\prime \Lambda(\vert \mathbf{x}-\mathbf{x}^\prime\vert)\frac{\partial^z}{\partial |\mathbf{x}^\prime | ^z}h(\mathbf{x}^\prime,t)+\eta(\mathbf{x},t)~,
\end{eqnarray}
where the scalar field $h(\mathbf{x},t)$ is parameterized by the coordinate
$\mathbf{x}$ and time $t$. The integral kernel $\Lambda(\vert \mathbf{x}-
\mathbf{x}^\prime\vert)$ of the spatial convolution integral represents the
generally non-local coupling of different sites $\mathbf{x}$ and $\mathbf{x}'$.
Moreover, $\partial^z/\partial|\mathbf{x}|^z$ is the multidimensional
Riesz-Feller fractional space derivative of order $z$ which is defined via its
Fourier transform through the functional relation \cite{samko}
\begin{equation}
\mathscr{F}\left\{\frac{\partial^z}{\partial|\mathbf{x}|^z}h(\mathbf{x},t)\right\}
\equiv-|\mathbf{q}|^z h(\mathbf{q},t).
\end{equation}
Here, $h(\mathbf{q},t)$ is the Fourier transform of $h(\mathbf{q},t)$. The
Gaussian noise $\eta(\mathbf{x},t)$ is fully determined by its first two
moments, $\langle \eta(\mathbf{x},t)\rangle=0$ and
\begin{equation}
\label{correlated noise}
\langle \eta(\mathbf{x},t)\eta(\mathbf{x}',t')\rangle=2k_BT \Lambda'(|\mathbf{x}
-\mathbf{x}'|)\delta(t-t'),
\end{equation}
where $\Lambda'(r)$ with $r=|\mathbf{x}-\mathbf{x}'|$ represents spatial
correlation properties of the noise. 

It is important to note that, in general, $\Lambda(r)\neq\Lambda'(r)$, that is,
both functions may be chosen independently. In what follows, to extract the
scaling properties of the GEM we first consider the general situation with long
ranged hydrodynamic-style interactions, $\Lambda(r)\sim r^{-\alpha_1}$, and
fractional Gaussian noise with long range spatial correlations, $\Lambda'(r)\sim
r^{-\alpha_2}$. We will discuss the following special cases:

(\textbf{a}) The interaction is local, $\Lambda(r)=\delta(r)$, and the noise is
an uncorrelated Gaussian random variable, $\Lambda^\prime(r)=\delta(r)$. This
special case corresponds to taking $\alpha_1=\alpha_2=d$.

(\textbf{b}) The interaction term $\Lambda$ is non-local with long-range
power-law interaction and the random noise $\eta$ has long-range correlations,
both with the same exponents $\alpha_1=\alpha_2=\alpha$.

(\textbf{c}) The interaction is local, $\Lambda(r)=\delta(r)$ ($\alpha_1=d$), and
the noise is long-range correlated, $\Lambda'(r)\propto r^{-\alpha}$ ($\alpha_2
=\alpha$).

(\textbf{d}) The interaction is non-local, $\Lambda(r)\propto r^{-\alpha}$
($\alpha_1=\alpha$), and the noise is uncorrelated and Gaussian, $\Lambda'(r)=
\delta(r)$ ($\alpha_2=d$).

In cases (\textbf{a}) and (\textbf{b}) the fluctuation-dissipation relation of
the second kind holds, whereas in cases (\textbf{c}) and (\textbf{d}) it is
violated. In the latter case the noise would then be viewed to be external. The
properties of the GEM in the presence of the fluctuation-dissipation theorem
have been studied analytically by Taloni \textbf{et al.}
\cite{{taloni2010,taloni2012,taloni2011,taloni2013}}. It is worthwhile
mentioning that $z=2$ in case (\textbf{a}) corresponds to the Edwards-Wilkinson
equation, and $z=4$ describes the universality class of the Mullins-Herring
equation \cite{surface1,surface2}. The Edwards-Wilkinson and Mullins-Herring
equations with long-range correlated power-law noise [cases (\textbf{a}) and
(\textbf{c})] were studied in Ref.~\cite{yu1994,pang1997,pang2010}. Krug
\textit{et al.} \cite{krug1997} used Eq.~(\ref{GEM}) with local interaction
$\Lambda(r)=\delta(r)$ to study the first passage statistics of locally
fluctuating interfaces. There Eq.~(\ref{GEM}) was solved numerically for the
special cases $z=2$ and $4$. Majumdar \textit{et al.} \cite{majumdar} considered
the same model to study the first-passage properties in space.  

Bearing in mind certain physical situations such as a cytoskeletal filament
pushing a single lipid in a vicinal membrane with some force \cite{taloni2011},
it will be interesting to consider the influence of such localized perturbations.
To that end we consider the extended GME equation
\begin{eqnarray}
\nonumber
\frac{\partial}{\partial t}{h}\left(\mathbf{x},t\right)&=&\int d^dx'\Lambda\left(
\mathbf{x}-\mathbf{x}'\right)\\
\nonumber
&&\hspace*{-1.2cm}\times\left[\frac{\partial^z}{\partial|\mathbf{x}'|^z}{h}(
\mathbf{x}',t)+\mathbf{F}\left\{h(\mathbf{x}',t),t\right\}\delta(\mathbf{x}'-
\mathbf{x}^{\star})\right]\\
&&\hspace*{-1.2cm}+\eta\left(\mathbf{x},t\right),
\label{GEM with local force}
\end{eqnarray}
such that the external force $\mathbf{F}$ acts only on the single (tagged) probe
at position $x^\star$ \cite{taloni2011}. This local force breaks the
translational invariance of the problem. We are interested in measuring the
average drift $\langle h(x^\star,t)\rangle_{F_0}$ in the perturbed system with
the constant force $\mathbf{F}\{ h(x',t),t\}=F_0\Theta(t)$ for different types
of the GEM. The forced problem will be discussed in Section
\ref{Chap: localizedpertGEM}.

\section{The GEM on a lattice}
\label{Chap:simulation}

To solve Eqs.~(\ref{GEM}) and (\ref{GEM with local force}) numerically, we
convert the dynamic formulation to discrete time and space in $d=1$. To that
end we define $t=n\Delta t$ with $n=1,2,\dots,N$ and $x=i\Delta x$ with $i=-
L/2,\dots,L/2$, where $\Delta t$ and $\Delta x$ are the grid steps in time
and space, respectively. To approximate the time derivative one can use a
simple forward Euler differential scheme, 
\begin{equation}
\frac{\partial h(x_i,t_n)}{\partial t}=\frac{h(x_i,t_{n+1})-h(x_i,t_n)}{
\Delta t}.
\end{equation}
In the following two Subsections, we review the methods to obtain a discrete
version of the fractional operator $\partial^z/\partial|x|^z$ and to generate
the correlated noise $\eta(x,t)$ with long-range correlation $\Lambda^\prime(r)
\sim r^{-\alpha}$. Then, we use the discrete version of Eqs.~(\ref{GEM}) and
(\ref{GEM with local force}) in our numerical simulations.

\subsection{The discretized fractional operator}

Rewriting the integral term of the GMEs (\ref{GEM}) and
(\ref{GEM with local force}) with a power-law kernel $\Lambda(r)$ in terms of
a space-fractional differential expression allows us to use known numerical
methods for analysis. Indeed the concept of fractional operators has been
successfully applied to a wide field of problems in physics, chemistry, finance,
biology and hydrology \cite{report,samko,podlubny,kilbas}. Here we employ the
discrete-space representation of the Riesz-Feller derivative $\partial^z/
\partial|x|^z$ of fractional order $z$. Different numerical methods have been
proposed to simulate such fractional operators \cite{yang}. We here pursue the
following approach. We rewrite the Riesz-Feller derivative in terms of the
standard Laplacian $\Delta^2$ as $\partial^z/\partial|x|^z:=-(-\Delta)^{z/2}$
\cite{saichev1997}, and then use the matrix transform method proposed by Ili\'c
\textit{et al} \cite{ilic2005} to approximate the discrete space fractional
operator (see also Appendix \ref{appendixA}).

Let us first consider the usual Laplacian in one dimension and a complete
set of orthogonal functions $\{\phi(x)\}$. In terms of the finite difference
method,
\begin{eqnarray}
\label{fourier laplace}
\Delta\phi(x)=\frac{\phi(x-a)-2\phi(x)+\phi(x+a)}{a^2},
\end{eqnarray}
where $a$ represents the lattice constant. With the Fourier representation
\begin{equation}
\phi(x)=\frac{1}{2\pi}\int\widehat{\phi}(q)e^{-iqx}dq,
\end{equation}
we obtain the Fourier transform of the discretized Laplacian
Eq.~(\ref{fourier laplace}) as
\begin{eqnarray}
\widehat{(\Delta)}\phi(q)=-[2-2\cos(qa)]\phi(q).
\end{eqnarray}
On the other hand one can find the elements of the matrix representation of the
Laplacian,
\begin{equation}
\mathbb{A}_{l,m}=-\int_0^{2\pi}\frac{dq}{2\pi}[2-2\cos(qa)]e^{iq(l-m)},
\end{equation}
where the tridiagonal matrix $\mathbb{A}\equiv\text{tridiag}(1,-2,1)$ has nonzero
elements only in the main diagonal and the first diagonals below and above the
main one.

We now use the approximation proposed by Ili\'c \textit{et al} (compare also
Appendix \ref{appendixA} and Refs.~\cite{ilic2005,yang}) to find the Fourier
representation of the fractional Laplacian. Namely, we start with the Fourier
representation of the discretized Laplacian $(-\Delta)$ with the minus sign,
$\lambda(q)=2[1-\cos(q)]$ and raise it to the appropriate power, $(2[1-\cos(q)])
^{z/2}$ \cite{zoia2007}. Here the lattice constant has been set equal to one.
The elements of the matrix $\mathbb{K}$, representing the discretized fractional
Laplacian $-(-\Delta)^{z/2}$ are then given by
\begin{eqnarray}
\mathbb{K}_{l,m}&=&-\int_0^{2\pi}\frac{dq}{2\pi}e^{iq(l-m)}\Big(2[1-\cos(q)]
\Big)^{z/2}\nonumber\\
&=&\frac{\Gamma(-z/2+n)\Gamma(z+1)}{\pi\Gamma(1+z/2+n)}\sin\left(\frac{z}{2}\pi
\right),
\label{HOLR} 
\end{eqnarray}
where $n=|l-m|$, and the fractional order $z\geq 1$. In the special case $z=2$
the $\mathbb{K}$ matrix is equal to the matrix $\mathbb{A}$ of the regular
Laplacian. Moreover, if $\alpha/2$ is an integer, then $\mathbb{K}(n)=(-1)^{
\alpha-n+1}C_{\alpha,n+\alpha/2}$ for $n\leq\alpha/2$ and $\mathbb{K}(n)=0$ for
$n>\alpha/2$, where the $C_{\alpha,n+\alpha/2}$ represent binomial coefficients
\cite{zoia2007}.

\subsection{Fractional Gaussian noise}

Several methods have been used to generate one-dimensional random processes
with long-range correlations, for instance, the successive random addition
method \cite{peitgen}, the Weierstrass-Mandelbrot function \cite{ausloos1985}
as well as the optimization method \cite{hamzehpour2006}. A very efficient way
to generate fractional Gaussian noise (fGn) is the modified Fourier filtering
(MFF) method \cite{makse1996}, compare also Appendix \ref{appendixB}.

Following Ref.~\cite{makse1996}, one needs a slightly modified correlation
function to deal with the singularity of $\Lambda'(r)$ at $r=0$ and to generate
the correlated noise. We use the form
\begin{eqnarray}
\label{nonsingLam}
\Lambda'_c(r)=(c^2+r^2)^{-\alpha/2},
\end{eqnarray}
with the asymptotically correct behavior $\Lambda'_c(r)\sim r^{-\alpha}$ at $r
\gg c$. The continuum limit of the spectral density $\Lambda'_c(q)$ becomes
\begin{eqnarray}
\label{modified powerlaw filter}
\Lambda'_c(q)\equiv\mathscr{F}\left\{\Lambda'_c (r)\right\}=\frac{\sqrt{\pi}2^{
1-\gamma}c^{-\gamma}}{\Gamma(\alpha/2)}|q|^{\gamma}K_{\gamma}(c|q|),
\end{eqnarray}
where $\gamma=(\alpha-1)/2$ and $K_\gamma$ is the modified Bessel function of
order $\gamma$. Then for small values of $c$ and $q$,
Eq.~(\ref{modified powerlaw filter}) leads to the asymptotic behavior $\Lambda'
_c(q)\sim q^{\alpha-1}$ (see Eq.~(\ref{FFM})).  

The numerical algorithm for generating correlated noise $\eta$ for arbitrary
values of $\alpha$ consists of the following steps:

(\textbf{i.}) Generate a one-dimensional array of uncorrelated Gaussian random
variables, $w_i$, and compute their Fourier transform $w_q$.

(\textbf{ii.}) Calculate $\{\eta(q,t_n)\}=\left[\Lambda_c'(q)\right]^{1/2}w_q$,
where $\Lambda_c'(q)$ is given by Eq.~(\ref{modified powerlaw filter}).

(\textbf{iii.}) Calculate the inverse Fourier transform $\eta(x_i,t_n)=
\mathscr{F}^{-1}\{\eta(q,t_n)\}$ to obtain the correlated noise with the desired
correlation exponent $\alpha$ in the real space.

We should note that we use periodic boundary condition, i.e., $\Lambda'_c(r)=
\Lambda'_c(r+L)$ in the interval $[-L/2,\dots,L/2]$, consequently we get the
correlated sample with the same periodicity. It is also possible to generate
a sample with natural boundary conditions. To this end, one first needs to
generate a sample with periodic boundary condition, of size $2L$, and then cuts
the sequences of the fGn time series into two separate parts with the same size
$L$, where each part obeys an open boundary condition.

For the uncorrelated case $\Lambda'(r)=\delta(r)$, the noise $\eta$ has a
Gaussian distribution and every $\eta(x_i,t_n)$ is an independent random
variable with zero mean and unit variance [with the convention $k_BT=1$ in
Eq.~(\ref{correlated noise})].

With these definitions we represent Eqs.~(\ref{GEM}) and (\ref{GEM with local
force}) in terms of discrete space and time variables $x_i=i\Delta x$ and $t_n
=n\Delta t$ in the form
\begin{eqnarray}
\label{discretGEM}
h_i^{n+1}&=&h_i^n+\Delta t\sum_{j=-L/2}^{L/2}\sum_{k=-L/2}^{L/2}\Lambda(|i-j|)
\mathbb{K}_{j,k}h_k^n\nonumber\\
&&+\sqrt{2\Delta t}\eta_i^n, 
\end{eqnarray}  
where $h_i^n$ approximates the field $h(x_i,t_n)$ at the $i$th lattice point and
the $n$th time step. At any given time step $n$, one needs to generate the random
process $\eta_i^n$ with the appropriate correlation function $\Lambda'$.
Analogously, the lattice version of the driven GEM with localized perturbation
becomes
\begin{eqnarray}
h_i^{n+1}&=&h_i^n+\Delta t\sum_{j=-L/2}^{L/2}\sum_{k=-L/2}^{L/2}\Lambda(|i-j|)
\mathbb{K}_{j,k}h_k^n\nonumber\\
&&+\Delta t\Lambda(|i-i^\star|)\mathbf{F}\{h(i^\star,n),n\}+\sqrt{2\Delta t}
\eta_i^n, 
\label{discretGEM perturb}
\end{eqnarray} 
where $i^\star$ corresponds to the position of the tagged probe. In the next
Section we present our numerical results and compare them with the analytical
predictions.

\section{Results}\label{Chap:Results}

To determine the time evolution of the scalar field $h(x,t)$ and to obtain the
dynamic scaling properties of the GEM, we simulated this model on a lattice of
size $L=4098$. All simulation measurements are based on an ensemble of $500$
realizations. In the simulations the time increment $\Delta t$ should be small
enough to ensure the stability of the numerical algorithm, and we find that
$\Delta t=0.05$ is a good working choice. As offset for $\Lambda'_c$ we choose
$c=0.05$. As already mentioned above, in order to avoid finite size effects
we impose periodic boundary conditions. At first we consider the unbiased
discrete GEM (\ref{discretGEM}) with non-thermal initial condition [$h(i,0)=0$,
$i\in[-L/2,\dots,L/2]$], and we measure the scaling exponents $\beta$ and $\tau(
q,\beta)$ of the second and $q$th order moments. Then we test the ergodic
properties of the GEM with non-thermal initial condition. Finally, we move to
the lattice version of the driven GEM (\ref{discretGEM perturb}) with localized
perturbation and measure the average drift for the tagged probe.

\subsection{Scaling laws and the $h$-correlation function}

The solution of the GEM (\ref{GEM}) has a continuous scale invariance property,
that is, for a physical observable $\mathcal{O}$ the relation
\begin{equation}
\mathcal{O}(\lambda\mathbf{x},\lambda^\nu t)=\mu(\lambda)\mathcal{O}(\mathbf{x},
t)
\end{equation}
arises, where $\mu(\lambda)$ is a power function of the scale factor $\lambda$.
This means that Eq.~(\ref{GEM}) does not change under a scaling transformation
$\mathbf{x}\rightarrow\lambda\mathbf{x}$ and $t\rightarrow\lambda^\nu t$,
together with the corresponding rescaling in the amplitude, $h\rightarrow
\lambda^\xi h$.

The scaling properties of the stochastic field $h(\mathbf{x},t)$ in a
$d$-dimensional space of linear size $L$ can be also characterized in
terms of the global interface width $W(t)$ defined by the root-mean-square
fluctuation of the random profile $h(x,t)$ at site $x$ and time $t$, that is,
\begin{equation}
W(t)=\left<\int dx[h(\mathbf{x},t)-\langle h(t)\rangle]^2\right>^{1/2},
\end{equation}
where $\langle h(t)\rangle=\int d^dxh(\mathbf{x},t)$. This width $W(t)$ scales as
\begin{equation}
\label{width}
W(t)\sim L^\xi f(t/L^\nu)\sim\left\{\begin{array}{ll}
t^\beta, & t\ll t_s\\[0.2cm] L^\xi, & t\gg t_s
\end{array}\right.,
\end{equation}
where $t_s=L^\nu$ is the so-called saturation time and $f(x)$ is a scaling
function with the property $f(x)\sim x^\beta$ for $x\ll 1$ and $f(x)\sim const.$
for $x\gg1$ \cite{surface1,surface2}. According to Eq.~(\ref{width}), we obtain
the constraint $\beta=\xi/\nu$ between the scaling exponents. With these 
relations we obtain the scaling exponents $\xi$, $\nu$, and $\beta$ for different
forms of the interaction kernel $\Lambda$ and the noise correlation function
$\Lambda'$. To this end we consider $\Lambda(r)\sim r^{-\alpha_1}$ and $\Lambda'(
r)\sim r^{-\alpha_2}$. If $\alpha_1=d$, the hydrodynamic interaction is local,
while $\alpha_2=d$ corresponds to a system with uncorrelated thermal noise. The
scale transformations $\mathbf{x}\rightarrow\lambda\mathbf{x}$ and $t\rightarrow
\lambda^\nu t$ transform the GEM (\ref{GEM}) according to
\begin{eqnarray}
\frac{\partial}{\partial t}h(\mathbf{x},t)&=&\lambda^{\nu-\gamma}\int d^dx'
\Lambda(\mathbf{x}-\mathbf{x}')\frac{\partial^z}{\partial|\mathbf{x}'|^z}h(
\mathbf{x}',t)\nonumber\\
&&+\lambda^{(\nu-\alpha_2)/2-\xi}\eta(\mathbf{x},t),
\end{eqnarray}
where $\gamma=z+\alpha_1-1$. The scale-invariance of the solution of the GEM
(\ref{GEM}) implies that $\nu=z+\alpha_1-d$ and $\xi=(z+\alpha_1-\alpha_2-d)/2$.
This specifies the dynamic scaling exponent
\begin{equation}
\label{generalbeta}
\beta=\frac{(z+\alpha_1-\alpha_2-d)}{2(z+\alpha_1-d)}.
\end{equation}

We now turn to determine the scaling properties of the $h$-correlation function
for the GEM with general interaction kernel $\Lambda(r)\sim r^{-\alpha_1}$ and
noise correlation $\Lambda'(r)\sim r^{-\alpha_2}$. Some previous measures of the
$h$-correlation for the special cases with $\alpha_1=\alpha_2=d$ [our case
(\textbf{a})] and $\alpha_1=\alpha_2=\alpha$ [our case (\textbf{b})] were
studied in Refs.~\cite{taloni2010,taloni2012,taloni2013}.

To derive the $h$-correlation function for the GEM with general interaction
kernel $\Lambda(r)$ and noise correlation $\Lambda'(r)$ we follow the method
put forward in Ref.~\cite{taloni2012}. We first consider the flat initial
condition $h(x,0)=0$, the so-called \textit{non-thermal initial condition\/}
\cite{taloni2012,krug1997}. We mention that the dynamics of the GEM depends on
the specific choice of the initial condition of Eq.~(\ref{GEM}), compare the
discussion in Ref.~\cite{taloni2012}. Then, the one-point, two-time correlation
function reads 
\begin{eqnarray}
\nonumber
\langle\delta_th(\mathbf{x},t)\delta_{t'}h(\mathbf{x},t')\rangle&=&\langle[h(
\mathbf{x},t)-h(\mathbf{x},0)]\\
\nonumber
&&\times[h(\mathbf{x},t')-h(\mathbf{x},0)]\rangle\\
&&\hspace*{-1.8cm}=K\left[(t+t')^{2\beta}-|t-t'|^{2\beta}\right],
\label{one-point two-time corr final nonthermal}
\end{eqnarray}
where $\beta$ matches the result of our above scaling arguments, compare
Eq.~(\ref{generalbeta}), and we find
\begin{equation}
K=\frac{2k_BT\pi^{d/2}}{(2\pi)^d\Gamma(d/2)}\frac{\Gamma(1-2\beta)}{z-d}\left(
\frac{(4\pi)^{d/2}}{2^\alpha}\frac{\Gamma((d-\alpha)/2)}{\Gamma(\alpha/2)}\right)
^{2\beta}.
\end{equation}
The dynamic scaling exponent $\beta$ for the different cases introduced in
Section \ref{Chap:Def} now takes assumes the values

(\textbf{a}) $\alpha_1=d$, $\alpha_2=d$, and $\beta=(z-d)/2z$.

(\textbf{b}) $\alpha_1=\alpha$, $\alpha_2=\alpha$, and $\beta=(z-d)/2(z+\alpha
-d)$.

(\textbf{c}) $\alpha_1=d$, $\alpha_2=\alpha$, and $\beta=(z-\alpha)/2z$.

(\textbf{d}) $\alpha_1=\alpha$, $\alpha_2=d$, and $\beta=(z+\alpha-2d)/2(z+
\alpha-d)$.

The results of our analysis for the two cases (\textbf{a}) and (\textbf{b}) are
in agreement with those of Refs.~\cite{taloni2012,taloni2013}, and our case
(\textbf{c}) agrees with the result of Ref.~\cite{krug1997}. 

It is worthwhile mentioning that the same calculations can be performed for the
system in the stationary state \cite{taloni2010}. The one-point, two-time
correlation can then be written as
\begin{eqnarray}
\nonumber
\langle\delta_th(\mathbf{x},t)\delta_{t'}h(\mathbf{x},t')\rangle_{\mathrm{st}}&&\\
&&\hspace*{-2cm}=K\left[(t)^{2\beta}+(t')^{2\beta}-|t-t'|^{2\beta}\right],
\label{one-point two-time corr final thermal}
\end{eqnarray}
where $\beta$ is again given by Eq.~(\ref{generalbeta}). Therefore, the dynamic
exponent $\beta$ is a universal quantity, that does not depend on the specific
initial condition. 

Note that in order to calculate the mean squared displacement $\langle(\delta h
)^2(t)\rangle$ and $\langle(\delta h)^2(t)\rangle_{\mathrm{st}}$ for the probe
particle, one should set $t=t'$ in Eqs.~(\ref{one-point two-time corr final
nonthermal}) and (\ref{one-point two-time corr final thermal}), respectively.
The mean squared displacement for these two cases follows in the forms
\begin{eqnarray}
\nonumber
\langle(\delta h)^2(t)\rangle&=&K(2t)^{2\beta},\\[0.2cm]
\langle(\delta h)^2(t)\rangle_{\mathrm{st}}&=&2Kt^{2\beta}.
\label{MSD}
\end{eqnarray}

\begin{figure}
\includegraphics[width=8.8cm]{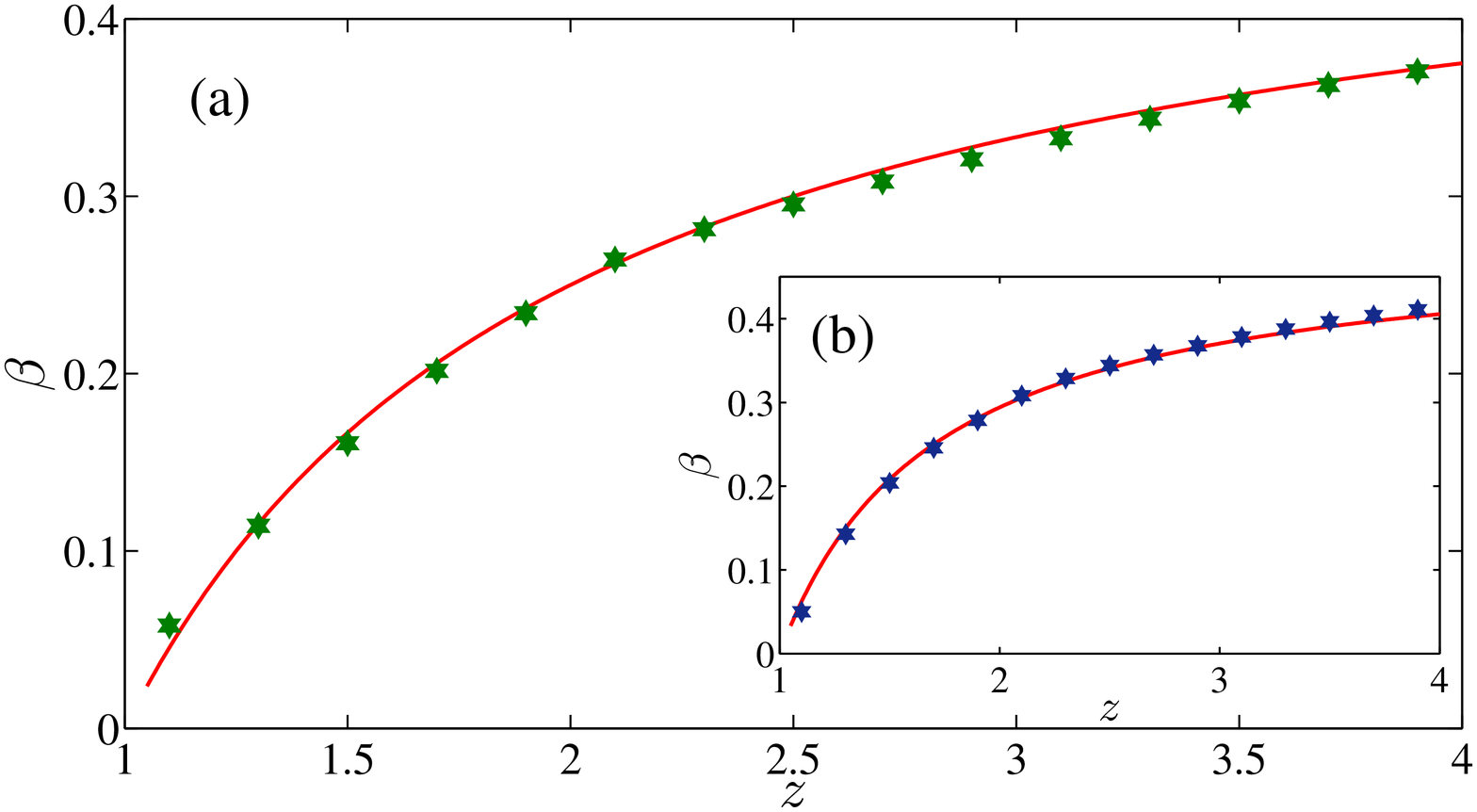}
\includegraphics[width=8.8cm]{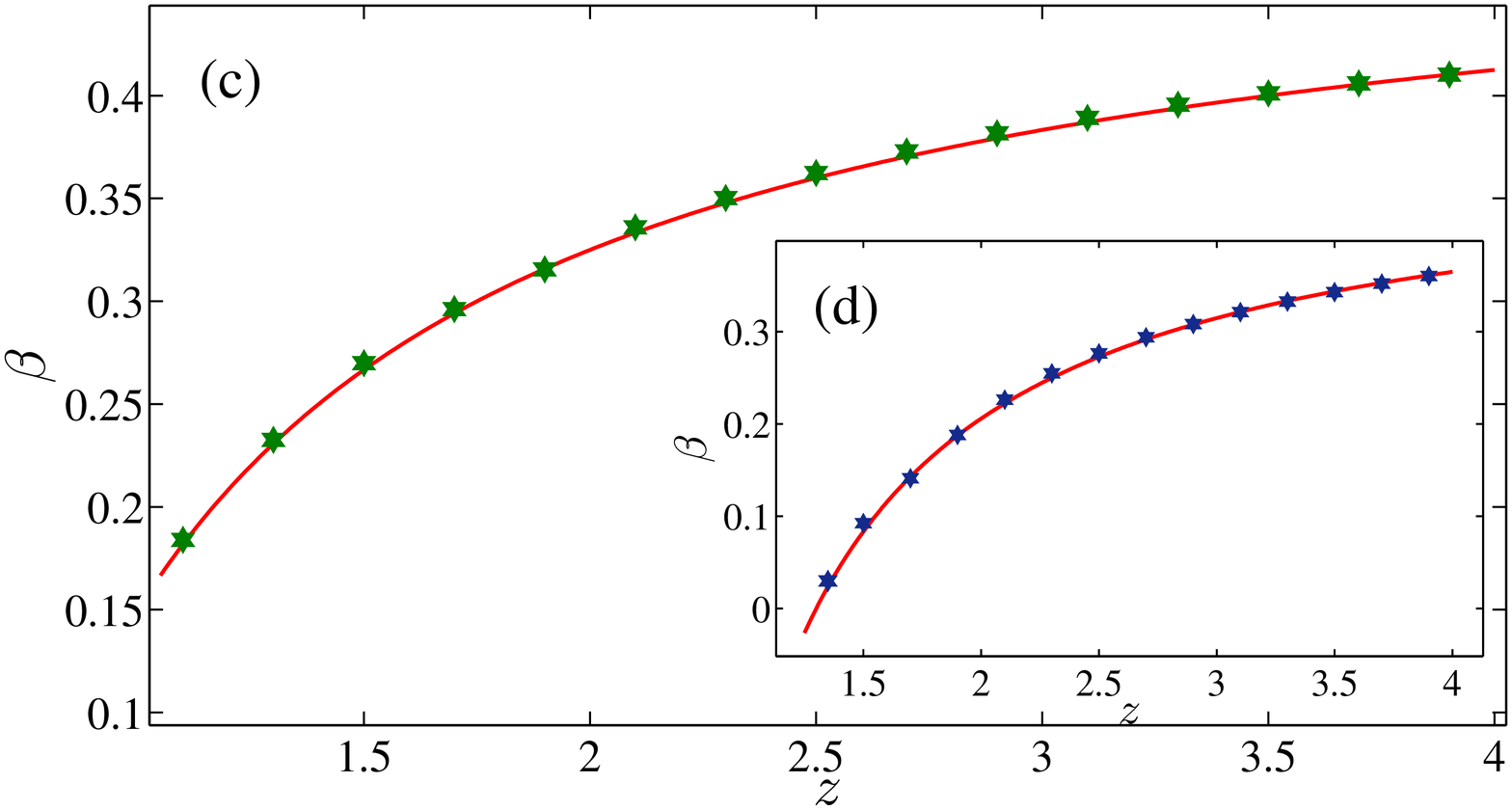}
\caption{(Color online) Comparison between theoretical predictions (solid curves)
and numerical results (symbols) for the subdiffusion exponent $\beta$ of the mean
squared displacement $\langle(\delta h)^2(t)\rangle$ of the probe particle. The
GEM (\ref{GEM}) with $d=1$ is specified by the interaction kernel $\Lambda(r)\sim
r^{-a_1}$ and correlated noise with $\Lambda'(r)\sim r^{-a_2}$, where
$\alpha_1$ and $\alpha_2$ for the different cases (\textbf{a}) to (\textbf{d})
depicted in panels (a) to (d) are chosen as (a) $\alpha_1=1$, $\alpha_2=1$; (b)
$\alpha_1=0.7$, $\alpha_2=0.7$; (c) $\alpha_1=1$, $\alpha_2=0.7$; and (d)
$\alpha_1=0.7$, $\alpha_2=1$.}
\label{Fig1}
\end{figure}

In Fig.~\ref{Fig1} we show numerical results for the subdiffusion exponent
$\beta$ as function of the fractional order $z$ for the cases (\textbf{a})
to (\textbf{d}) introduced in Section \ref{Chap:Def}. The exponent is measured
from the power-law dependence of the mean squared displacement with time, see
the first equality in Eq.~(\ref{MSD}). The results of the numerical simulations
are shown by the symbols, and the solid curves demonstrate the analytical result
(\ref{generalbeta}). We observe excellent agreement with the theoretical result
for all our cases in the interesting range for $z$ between 1 and 4.

\subsection{Scaling properties of $q$th order moments}

We now turn to the scaling properties of the $q$th order moments $\langle|\delta
h(t)|^q\rangle$. According to the scale-invariance property,
\begin{equation}
h(s^{1/\nu}\mathbf{x},st)\equiv s^{\xi/\nu}h(\mathbf{x},t), \,\,\, s>0 , 
\end{equation} 
and the condition $\langle|\delta h(st)|^q\rangle=s^{q\xi/\nu}\langle|\delta h(t)
|^q\rangle$, we find
\begin{eqnarray}
\langle|\delta h(t)|^q\rangle\sim t^{\tau(q,\beta)},
\end{eqnarray}
where $\beta=\xi/\nu$ and $\tau(q,\beta)=q\beta$. When the exponent $\tau(q,
\beta)$ is a linear function of $q$, the process is referred to as a mono-scale
process, and the stochastic profile $h(\mathbf{x},t)$ is non-intermittent
\cite{taqqu}. 

\begin{figure}
\includegraphics[width=8.8cm]{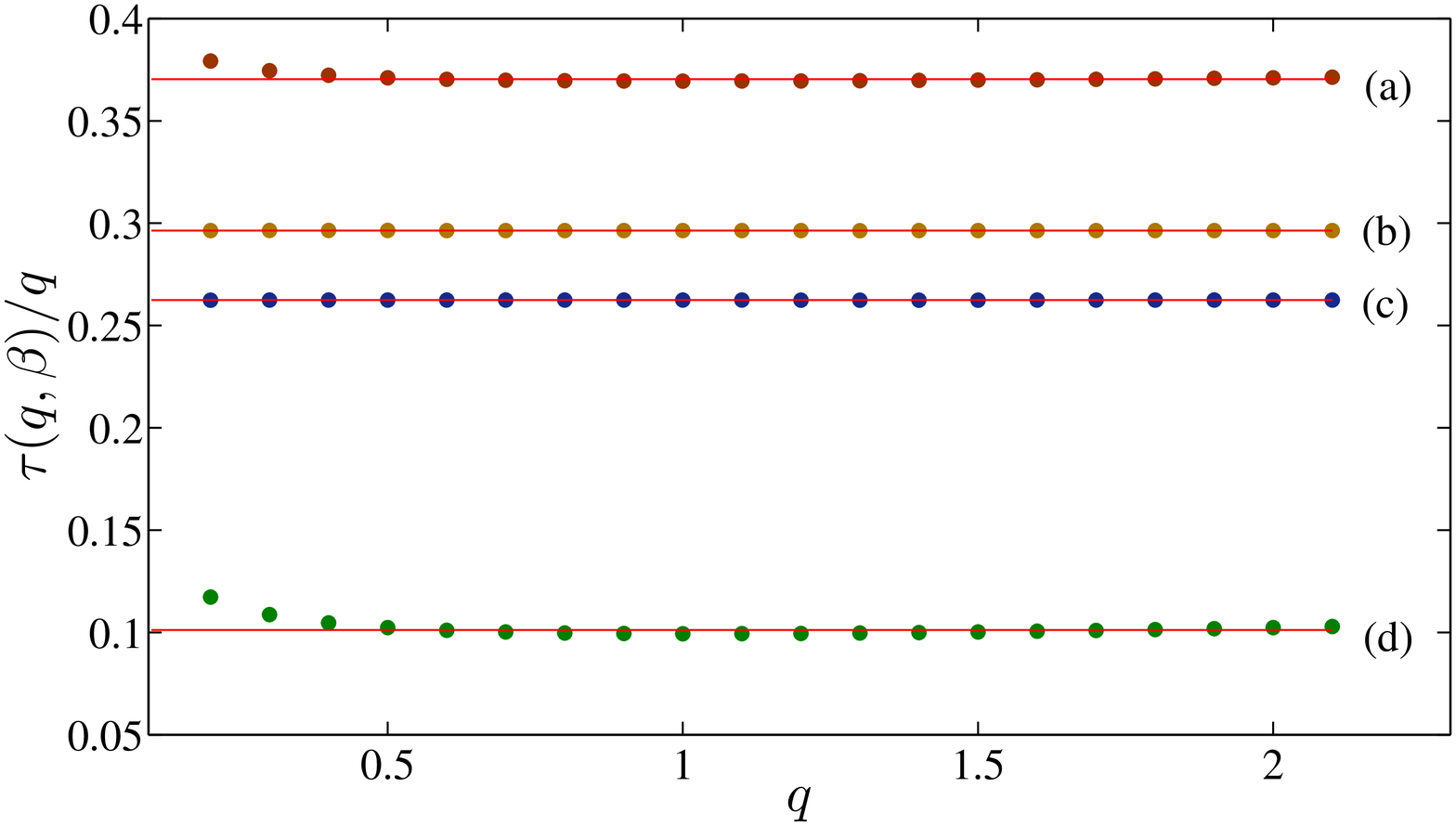}
\caption{(Color online) The dots show the values of $q^{-1}\tau(q,\beta)$
obtained in numerical simulations of the GEM with interaction kernel
$\Lambda(r)\sim r^{-\alpha_1}$ and correlated noise with $\Lambda^\prime(r)\sim
r^{-\alpha_2}$. The parameters $\alpha_1$, $\alpha_2$ and $z$ correspond to
the four cases (\textit{a})-(\textit{d}) and are equal to: (a) $\alpha_1 =
1$, $\alpha_2 = 1$, $z = 4.0$ (b) $\alpha_1 =0.7$, $\alpha_2=0.7$, $z =
2.0$ (c) $\alpha_1 = 1$, $\alpha_2 = 0.7$, $z = 1.3$ (d)$\alpha_1=0.7$,
$\alpha_2 = 1$, $z = 1.5$. Solid lines show the corresponding values of the
exponent $\beta$. This plot demonstrates the mono-scaling (non-intermittent)
behavior of fluctuations.}
\label{Fig2}
\end{figure}

We studied the scaling behavior of the $q$-th moment numerically.
In Fig. (\ref{Fig2}) $\tau(q,\beta)/q$ is plotted vs $q$ for the four
paradigmatic examples, the cases (\textit{a})-(\textit{d}). The figure
shows that the $q^{-1}\tau(q,\beta)$ is equal to $\beta$ and independent
of $q$, which demonstrates that the height fluctuations in the GEM are not
intermittent.

\subsection{Ergodic properties}

In the two previous Subsections we obtained the scaling exponents $\beta$
and $\tau(q,\beta)$ for the GEM with non-thermal initial condition. For this
purpose we used the ensemble average of the second and $q$th order moments.
For example, to determine the subdiffusion exponent $\beta$ one needs to obtain
the ensemble average of the observable $(\delta h)^2$. In many experiments,
however, one measures time averages of physical observables (see, for instance,
Refs.~\cite{He2008,burov2011}). For an ergodic process, the long time average of
an observable produces the same result as the corresponding ensemble average,
while for a non-ergodic process the correct interpretation of the time average
requires a separate theory. We here consider a single trajectory of length $T$
(measurement time) and define the time average as
\begin{equation}
\label{ensemble aver}
\overline{\delta^2 h(\Delta)}=\frac{1}{T-\Delta}\int_0^{T-\Delta}dt\left[h(x,
t+\Delta)-h(x,t)\right]^2,
\end{equation}  
where $\Delta$ denotes the lag time. It was shown in Ref.~\cite{taloni2012}
that the additional ensemble average $\langle\overline{\delta^2 h}\rangle$ of
the quantity (\ref{ensemble aver}) for systems with non-thermal initial
condition tends to the value of the ensemble averaged MSD $\langle(\delta h)^2
(t)\rangle_{\mathrm{st}}$ in the stationary state, if $\Delta/T\rightarrow 0$.
This means that the process is ergodic, and sufficiently long time averages
reproduce the exact behavior predicted by the ensemble quantities.

A useful quantity to measure the fluctuations between different realizations of
a dynamic process is the probability density function of the amplitude scatter
$\phi(\varepsilon)$ in terms of the dimensionless variable $\varepsilon=\overline{
\delta^2h}/\langle\overline{\delta^2h}\rangle$ \cite{He2008,burov2011,jae}. Thus,
$\phi(\varepsilon)$ measures how reproducible individual realizations $\overline{
\delta^2h}$ are with respect to the ensemble mean of the time averages, $\langle
\overline{\delta^2h}\rangle$. For an ergodic system, $\phi(\epsilon)$ has bell
shape around the ergodic value $\epsilon=1$, and for long measurement times $T$
it converges to a $\delta$-peak, $\lim_{T\to\infty}\phi(\epsilon)=\delta(\epsilon
-1)$ \cite{He2008,jae}.

Another measure of ergodic violation is the ergodicity breaking parameter 
\cite{He2008}
\begin{equation}
\label{EBP}
\mathrm{EB}=\lim_{T\rightarrow\infty}\frac{\left<\left(\overline{\delta^2h}
\right)^2\right>-\left<\overline{\delta^2h}\right>^2}{\left<\overline{\delta^2
h}\right>^2}.
\end{equation}
The sufficient condition for ergodicity is $\mathrm{EB}=0$.

\begin{figure}
\includegraphics[width=9cm]{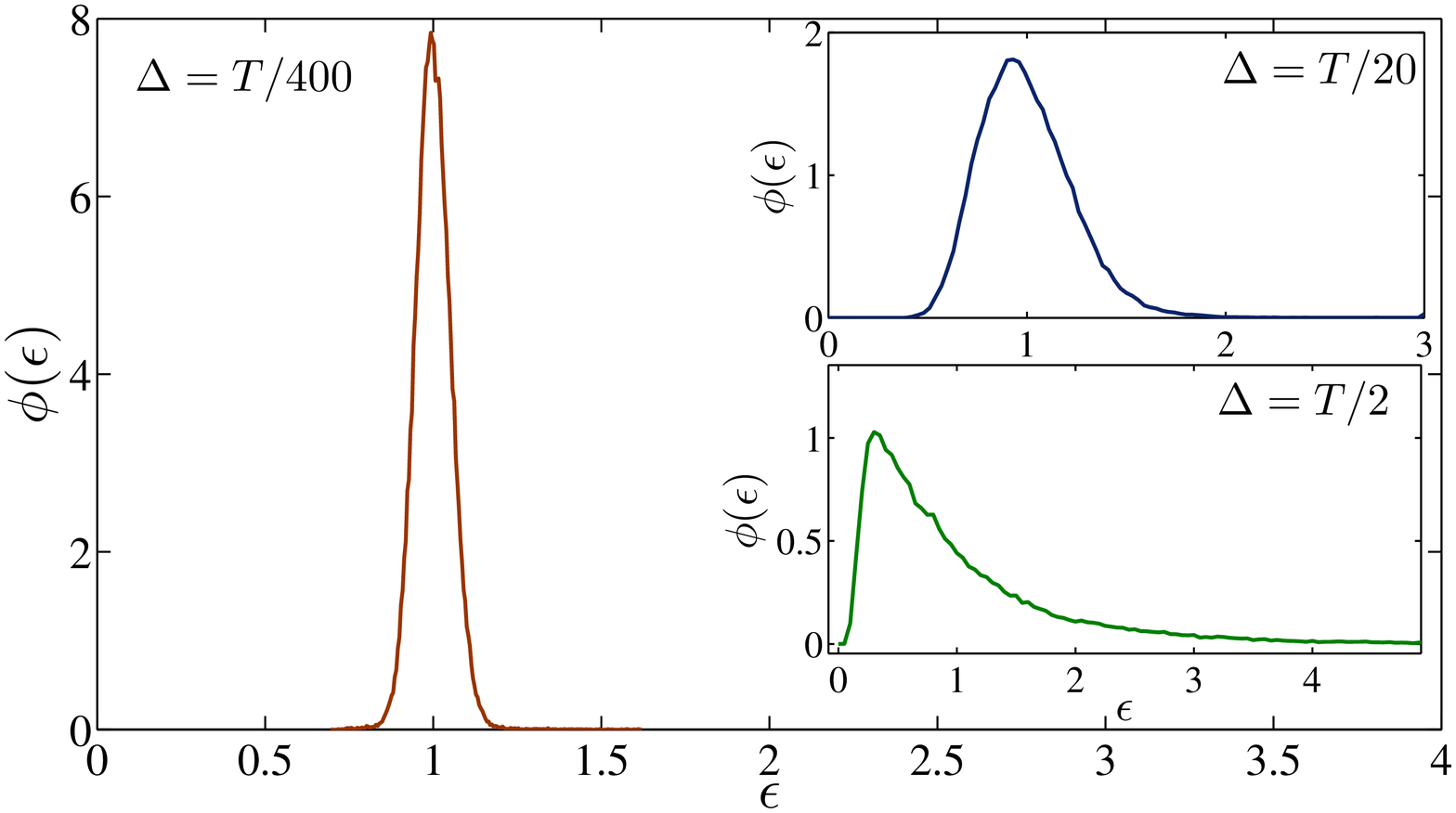}
\includegraphics[width=9cm]{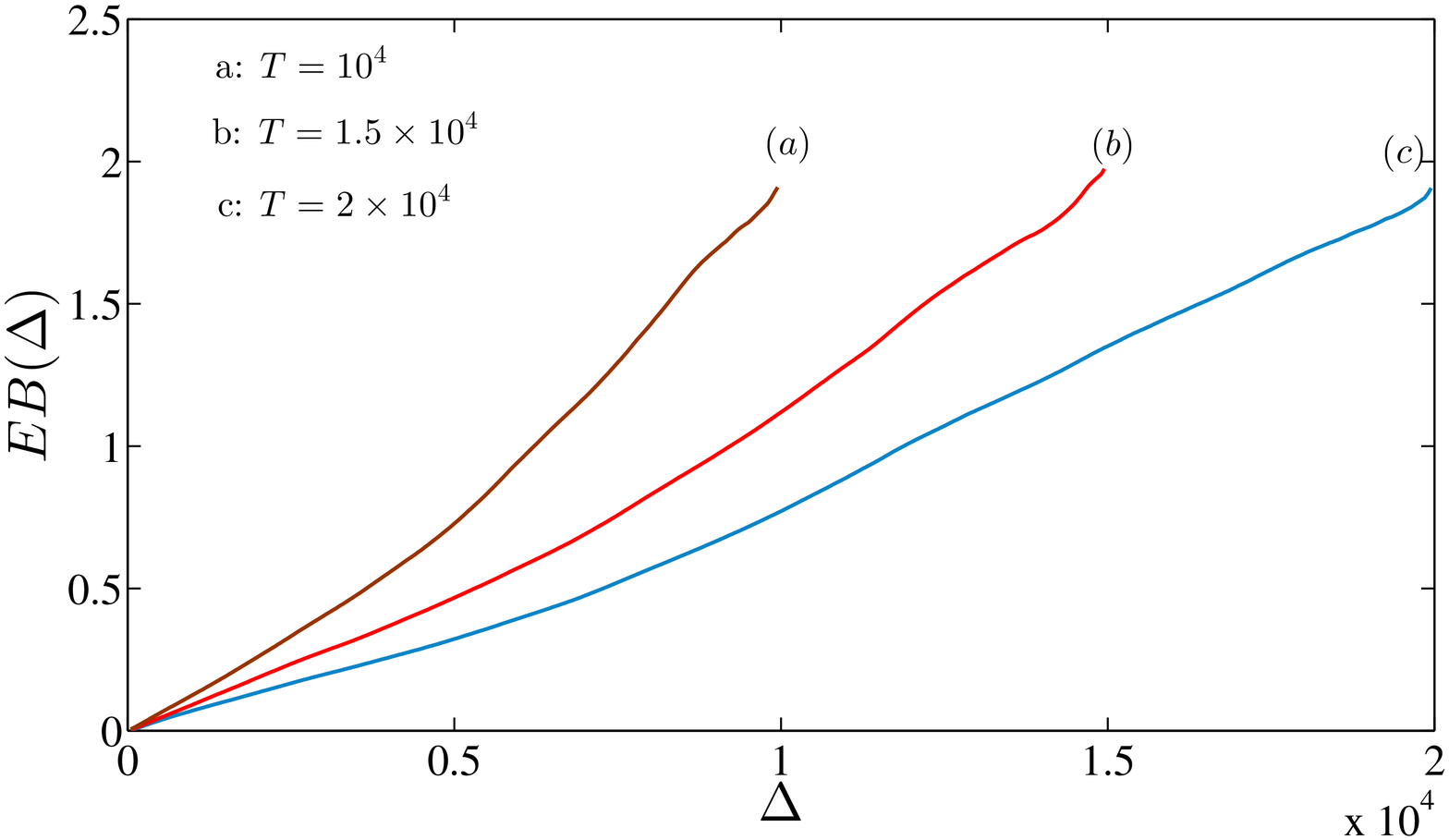}
\caption{(Color online) Top panel: Probability density function of the scaled
random variable $\epsilon=\overline{\delta^2 h}/\langle\overline{\delta^2 h}
\rangle$ for different values of the lag time $\Delta$. The measurement time
$T=2\times10^4$. Bottom panel: Ergodicity breaking parameter $\mathrm{EB}$ as
a function of the lag time $\Delta$ for three different values of $T$.}
\label{Fig4}
\end{figure}

Here we restrict ourselves to the special case (\textbf{b}). To study the ergodic
properties of the GEM, we calculate the amplitude scatter PDF $\phi(\epsilon)$
and the ergodicity breaking parameter $\mathrm{EB}$. In the top panel of
Fig.~\ref{Fig4} we show that the shape of $\phi(\epsilon)$ becomes sharper
with decreasing $\Delta$ when the measurement time $T$ is fixed. In the limit
$\Delta\ll T$, the PDF has a peak close to the ergodic value $\overline{\delta
^2h}/\langle\overline{\delta^2h}\rangle=1$, which indicates the ergodicity of
the process. In the bottom panel of Fig.~\ref{Fig4} we depict the ergodicity
breaking parameter $\mathrm{EB}$ as a function of the lag time $\Delta$ for
different values of the measurement time $T$. We see that indeed the ergodicity
breaking parameter converges to the ergodic value $\mathrm{EB}\rightarrow0$ for
$\Delta/T\rightarrow0$.

\subsection{The GEM with localized perturbation}
\label{Chap: localizedpertGEM}

In the preceding Section \ref{Chap:Results} we studied the properties of the
unbiased GEM. We now report results of numerical simulations of the driven GEM
with a constant localized perturbation, $\mathbf{F}\{h(\mathbf{x}',t),t\}=F_0
\Theta(t)$, compare Eq.~(\ref{GEM with local force}). We consider the motion of
a tagged probe located at $x^\star$. The results for an untagged probe will be
presented elsewhere. Obviously, the stochastic term in Eq.~(\ref{GEM with local
force}) does not make a contribution to the average drift $\langle h(\mathbf{x}^
\star,t)\rangle_{F_0}$, since $\langle\eta(x,t)\rangle=0$. Thus, basically, the
average drift is determined by the nature of the hydrodynamic friction kernel
$\Lambda(r)$. Following Refs.~\cite{taloni2011,taloni2013} we determine the
average drift,
\begin{equation}
\label{scaling tagged prob}
\langle h(\mathbf{x}^\star,t)\rangle_{F_0}\propto F_0t^{2\beta'},
\end{equation}
where the dynamic scaling exponent is $\beta'=(z-d)/2z$ for the local and
$\beta'=(z-d)/2(z+\alpha-d)$ for the non-local hydrodynamic interaction, where
the former expression holds for the cases (\textbf{a}) and (\textbf{c}), while
the latter formula is valid for the cases (\textbf{b}) and (\textbf{d}). Note
that $\beta'=\beta$ for the GEM obeying the fluctuation-dissipation relation of
the second kind, corresponding to local hydrodynamic interaction and uncorrelated
noise [case (\textbf{a})] and that with non-local interaction and correlated
noise [case (\textbf{b})]. Thus, the Einstein relation 
\begin{equation}
\label{einstein relation}
\langle h(\mathbf{x}^\star,t)\rangle_{F_0}=\frac{\langle(\delta h)^2(t)\rangle_{
\mathrm{st}}}{2k_BT}F_0
\end{equation}
holds for the tagged probe in the two cases (\textbf{a}) and (\textbf{b}), where
$\langle(\delta h)^2(t)\rangle_{\mathrm{st}}$ is defined  by Eq. (\ref{MSD}). 
  
\begin{figure}
\includegraphics[width=9cm]{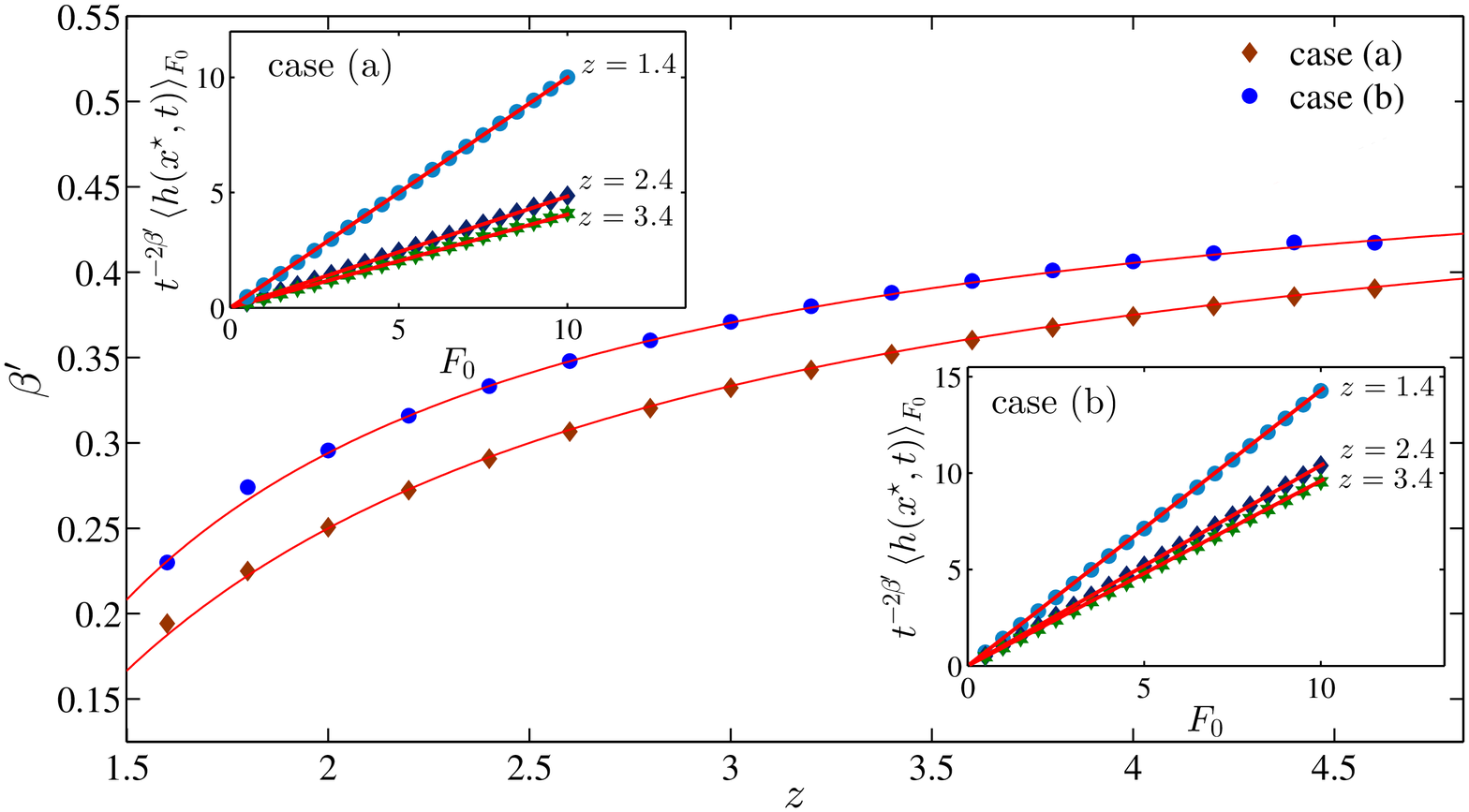}
\caption{(Color online) Dynamic scaling exponent $\beta'$ for the GEM with local
[cases (\textbf{a}) and (\textbf{c})] and nonlocal [cases (\textbf{b}) and
(\textbf{d}), with $\alpha=0.7$] hydrodynamic interaction. Main panel: numerical
results are shown by the symbols, whereas the solid (red) lines correspond to
the theoretical relations $\beta'=(z-1)/2z$ and $(z-1)/2(z+\alpha-1)$ for the
two cases (\textbf{a}) and (\textbf{b}), respectively. Insets: The relation
between $\langle h(x^\star,t)\rangle_{F_0}$ and $F_0$ for the tagged tracer in
the GEM [cases (\textbf{a}) and (\textbf{b})] for different fractional order
$z$. Solid red lines correspond to the analytical expression $t^{-2\beta'}
\langle h(x^\star,t)\rangle_{F_0}=\frac{K}{k_B T}F_0$ [see
Eqs.~(\ref{scaling tagged prob}) and (\ref{einstein relation})].}
\label{Fig5}
\end{figure}

We simulated the GEM with constant local force on a one-dimensional lattice, see
Eq.~(\ref{discretGEM perturb}). Then we calculate the average drift and extract
the dynamic exponent $\beta'$ according to Eq.~(\ref{scaling tagged prob}). The
results are shown in Fig.~\ref{Fig5}. The main panel depicts $\beta'$ as a
function of $z$ for local [case (\textbf{a})] and non-local [case (\textbf{b}]
interactions. The results of the simulations shown by the symbols perfectly
agree with the analytical findings (solid curves), i.e., $\beta'=(z-1)/2z$ for
the local and $\beta'=(z-1)/2(z+\alpha-1)$ for the non-local case, respectively.
In addition, in the insets we show $\langle h(\mathbf{x}^\star,t)\rangle_{F_0}$
as a function of the applied force $F_0$ for several values of $z$.

\section{Conclusions and Discussions}\label{Chap:Conclusions}

In this paper, we studied the interface dynamics of the generalized elastic
model with two types of interactions, local and long-range non-local ones, in
the presence of uncorrelated as well as long-range correlated noise. We
generalized some of the previous results from Taloni \textit{et al.}
\cite{taloni2010,taloni2012,taloni2011} on the scaling properties of the
GEM, and we developed a discrete numerical scheme to simulate the one-dimensional
GEM on the lattice by using a discretized version of the Riesz-Feller fractional
operator. We performed numerical simulations and measured the dynamics scaling
exponents $\beta$ for the second moment and $\tau(q,\beta)$ for the $q$th order
moments of the random field $h$ for four paradigmatic cases of the GEM. We also
analyzed the ergodic properties of the GEM and demonstrated the ergodicity of
the process by measuring the amplitude scatter of individual trajectories and
the ergodicity breaking parameter. Finally, we simulated the driven GEM with
localized perturbation and measured the scaling exponent $\beta'$ from the
scaling properties of the mean drift of a tagged probe. All numerical results
are in perfect agreement with the analytics, thus supporting the numerical
scheme developed herein.

It will be interesting to apply this numerical algorithm to other relevant
aspects of interface dynamics, such as the presence of quenched disorder,
and GEM with nonlinear terms. Another direction is to develop numerical
tools to study the GEM in higher dimensions.

\begin{acknowledgments}
The authors are indebted to A. Taloni for helpful discussions. MGN thanks S.
Rouhani and V. Palyulin for support and motivating discussions. MGN acknowledges
financial support from University of Potsdam. RM is supported by the Academy of
Finland (FiDiPro scheme).
\end{acknowledgments}

\begin{appendix}

\section{The matrix transform method}
\label{appendixA}

The discrete space fractional operator can be efficiently generated by the
matrix transform algorithm proposed in Refs.~\cite{yang,ilic2005}. This
method is based on the following definition \cite{ilic2005}: Consider the
Laplacian $(-\Delta)$ on a bounded region $\mathfrak{D}$, with a complete
set of orthonormal eigenfunctions $\phi_n$ and eigenvalues $\lambda^2_n$,
i.e., $(-\Delta)\phi_n=\lambda^{2}_n\phi_n$. An orthogonal and complete set
of functions $\{\phi_n\}$ may be used to expand an arbitrary function
$f(\textbf{x})$ in the following form
\begin{eqnarray}
f=\sum_{n=1}^\infty c_n\phi_n, \quad \text{with } c_m=\int_{\mathfrak{D}}
\phi_n(\textbf{x})f(\textbf{x})d\textbf{x}.
\end{eqnarray}
Then, for any $f$ one can define $(-\Delta)^{z/2}$ as
\begin{equation}
\label{app:2}
(-\Delta)^{z/2}f=\sum_{n=1}^\infty c_n\lambda^z_n\phi_n.
\end{equation}  
It is worthwhile mentioning that the complete set of functions $\{\phi_n\}$ is
also the eigensolution of the fractional operator, $(-\Delta)^{z/2}\phi_n=
\lambda^z_n\phi_n$. This definition provides a new method and corresponding
numerical scheme to approximate a space-fractional operator.

\section{The modified Fourier filtering method}
\label{appendixB}

The celebrated fractional Gaussian noise can be efficiently generated by the
algorithm proposed in Ref.~\cite{makse1996}. Consider the noise $\eta(x,t)$
with the correlation function 
\begin{eqnarray}
\langle\eta(x,t)\eta(x',t')\rangle\propto\Lambda'(r)\delta{(t-t')},
\end{eqnarray}
where $r=|x-x'|$ and $\Lambda'(r)\sim r^{-\alpha}$ in the limit $r\rightarrow
\infty$. For the fixed time instant $t_n$, the noise is generated on a uniform,
one-dimensional grid with $L$ points. Following the discrete Fourier
transformation, the Fourier component of the correlated noise $\eta(x_i,t_n)$
is defined by
\begin{eqnarray}
\eta(q,t_n)=\sum_{x_i=-L/2}^{L/2}\eta(x_i,t_n)e^{-iqx_n},
\end{eqnarray}
where $q$ assumes the values $q=2\pi m/L$ with $m=\{-L/2,\dots,L/2\}$. The
idea of the Fourier filtering method is to simulate a process $\eta(q,t_n)$
with the spectral density 
\begin{equation}
\label{powelaw filter}
\Lambda'(q)=\langle\eta(q,t_n)\eta(-q,t_n)\rangle\sim q^{\alpha-1},
\end{equation}
for $q\rightarrow0$, and transform the resulting series to real space. The
correlated noise is then constructed by filtering the Fourier components of a
sequence of normally distributed random numbers $\{w_i\}_{i=1,\dots,L}$ with
the correlation function $\langle w_iw_j\rangle\sim\delta_{i,j}$ and the
Fourier transform $w_q$. Then one generates the Fourier transform coefficients
of the correlated noise by
\begin{eqnarray}
\label{FFM}
\eta(q,t_n)=\left[\Lambda'(q)\right]^{1/2}w_q.
\end{eqnarray}

\end{appendix}


\begin{thebibliography}{99}

\bibitem{surface1} P. Meakin, \textit{Fractals, Scaling and Growth far from
Equilibrium} (Cambridge University Press, Cambridge, UK, 1998).

\bibitem{surface2} A. L. Barab\'si  and H. E. Stanley, \textit{Fractal Concepts
in Surface Growth} (Cambridge University Press, Cambridge, UK, 1995)

\bibitem{granek1997} R. Granek, J. de Physique II \textbf{7}, 1761 (1997);
R. Granek and J. Klafter, Europhys. Lett. \textbf{56}, 15 (2007).

\bibitem{single} L. Lizana, T. Ambj{\"o}rnsson, A. Taloni, E. Barkai, and M. A.
Lomholt, Phys. Rev. E \textbf{81}, 051118 (2010); M. A. Lomholt, L. Lizana, and
T. Ambj{\"o}rnsson, J. Chem. Phys. \textbf{134}, 045101 (2011).

\bibitem{zilman2002} A. G. Zilman and R. Granek, Chem. Phys.
\textbf{284}, 195 (2002).

\bibitem{helfer2000} E. Helfer, S. Harlepp, L. Bourdieu, J. Robert,
F. MacKintosh, and D. Chatenay, Phys. Rev. Lett. \textbf{85}, 457 (2000).

\bibitem{doi} M. Doi and S. F. Edwards, \textit{The Theory of Polymer
Dynamics} (Clarendon Press, Oxford, UK, 1986).

\bibitem{amblard1996} F. Amblard, A. C. Maggs, B. Yurke, A. N.
Pargellis, and S. Leibler, Phys. Rev. Lett. \textbf{77}, 4470 (1996).

\bibitem{everaers1999} R. Everaers, F. J{\"u}licher, A. Ajdari, and A.
Maggs, Phys. Rev. Lett. \textbf{82}, 3717 (1999).

\bibitem{caspi1998} A. Caspi, M. Elbaum, R. Granek, A. Lachish, and
D. Zbaida, Phys. Rev. Lett. \textbf{80}, 1106 (1998).

\bibitem{buldyrev1992} S. Buldyrev, A.-L. Barab{\'a}si, F. Caserta,
S. Havlin, H. Stanley, and T. Vicsek, Phys. Rev. A \textbf{45}, 8313 (1992).

\bibitem{nissen2001} J. Nissen, K. Jacobs, and J. O. R{\"a}dler,
Phys. Rev. Lett. \textbf{86}, 1904 (2001).

\bibitem{bustingorry2006} S. Bustingorry, L. F. Cugliandolo, and D. Dominguez,
Phys. Rev. Lett. \textbf{96}, 27001 (2006); U. Dobramysl, H. Assi, M. Pleimling,
and U. C. T{\"a}uber, arXiv preprint arXiv:1211.6929 (2012).

\bibitem{bru1998} A. Br{\'u}, J. M. Pastor, I. Fernaud, I. Br{\'u}, S.
Melle, and C. Berenguer, Phys. Rev. Lett. \textbf{81}, 4008 (1998).

\bibitem{podlubny} I. Podlubny, \textit{Fractional Differential Equations}
(Academic Press, London, UK, 1999).

\bibitem{vankampen} N. G. van Kampen, \emph{Stochastic processes in chemistry
and physics} (North Holland, Amsterdam, NL, 1981).

\bibitem{taloni2010} A. Taloni, A. Chechkin, and J. Klafter, Phys.
Rev. Lett. \textbf{104}, 160602 (2010); Rev. E \textbf{82}, 061104 (2010).

\bibitem{taloni2012} A. Taloni, A. Chechkin, and J. Klafter, Europhys.
Lett. \textbf{97}, 30001 (2012).

\bibitem{taloni2011} A. Taloni, A. Chechkin, and J. Klafter, Phys. Rev.
E \textbf{84}, 021101 (2011).

\bibitem{taloni2013} A. Taloni, A. Chechkin, and J. Klafter,
Mathematic. Modeling Nat. Phenomena \textbf{8}, 127 (2013).

\bibitem{REM} In the present work the exponent $\beta$ corresponds to $\beta/2$
from Taloni \textit{et al.} \cite{taloni2010,taloni2012,taloni2011,taloni2013}.

\bibitem{struik1978} L. C. E. Struik, \emph{Physical aging in amorphous
polymers and other materials}, vol. 106 (Elsevier Amsterdam, The Netherlands,
1978).

\bibitem{bouchaud} J.-P. Bouchaud and A. Georges, Phys. Rep. \textbf{195}, 127
(1990).

\bibitem{report} R. Metzler and J. Klafter, Phys. Rep. \textbf{339}, 1 (2000);
J. Phys. A \textbf{37}, R161 (2004).

\bibitem{bouchaud1992} J. P. Bouchaud, J. de Physique I \textbf{2}, 1705
(1992).

\bibitem{cugliandolo1995} L. Cugliandolo and J. Kurchan, Philos. Mag. B
\textbf{71}, 501 (1995).

\bibitem{monthus1996} C. Monthus and J.-P. Bouchaud, J. Phys. A
\textbf{29}, 3847 (1996).

\bibitem{rinn2000} B. Rinn, P. Maass, and J.-P. Bouchaud, Phys. Rev.
Lett. \textbf{84}, 5403 (2000).

\bibitem{barkai2003} E. Barkai, Phys. Rev. Lett. \textbf{90}, 104101 (2003).

\bibitem{bel2005} G. Bel and E. Barkai, Phys. Rev. Lett. \textbf{94}, 240602
(2005).

\bibitem{rebenshtok2007} A. Rebenshtok and E. Barkai, Phys. Rev.
Lett. \textbf{99}, 210601 (2007); J. Stat. Phys. \textbf{133}, 565 (2008).

\bibitem{johannes} J. H. P. Schulz, E. Barkai, and R. Metzler, Phys. Rev. Lett.
\textbf{110}, 020602 (2013).

\bibitem{He2008} Y. He, S. Burov, R. Metzler, and E. Barkai, Phys. Rev.
Lett. \textbf{101}, 058101 (2008); A. Lubelski, I. M. Sokolov, and J. Klafter,
Phys. Rev. Lett. \textbf{100}, 250602 (2008).

\bibitem{burov2011} S. Burov, J.-H. Jeon, R. Metzler, and E. Barkai,
Phys. Chem. Chem. Phys. \textbf{13}, 1800 (2011); J.-H. Jeon, E. Barkai, and
R. Metzler, J. Chem. Phys. \textbf{139}, 121916 (2013); E. Barkai, Y. Garini,
and R. Metzler, Phys. Today \textbf{65}(8), 29 (2012).

\bibitem{Joen2011} S. M. A. Tabei, S. Burov, H. Y. Kim, A. Kuznetsov, T. Huynh,
J. Jureller, L. H. Philipson, A. R. Dinner, and N. F. Scherer, Proc. Natl.
Acad. Sci. USA \textbf{110}, 4911 (2013); A. V. Weigel, B. Simon, M. M. Tamkun,
and D. Krapf, \emph{ibid.} \textbf{108}, 6438 (2011); J.-H. Jeon, V. Tejedor,
S. Burov, E. Barkai, C. Selhuber-Unkel, K. Berg-S{\o}rensen, L. Oddershede,
and R. Metzler, Phys. Rev. Lett. \textbf{106}, 048103 (2011); I. Y. Wong, M. L.
Gardel, D. R. Reichman, E. R. Weeks, M. T. Valentine, A. R. Bausch, and D. A.
Weitz, \emph{ibid.} \textbf{92}, 178101 (2004); Q. Xu, L. Feng, R. Sha, N. C.
Seeman, and P. M. Chaikin, \emph{ibid.} \textbf{106}, 228102 (2011).

\bibitem{cherstvy2013} A. G. Cherstvy, A. V. Chechkin, and R. Metzler,
New J. Phys. \textbf{15}, 083039 (2013); A. G. Cherstvy and R. Metzler,
E-print arXiv:1307.6407.

\bibitem{lomholt} M. A. Lomholt, L. Lizana, R. Metzler, and T. Ambj{\"o}rnsson,
Phys. Rev. Lett. \textbf{110}, 208301 (2013).

\bibitem{deng2009} W. Deng and E. Barkai, Phys. Rev. E \textbf{79},
011112 (2009); J.-H. Jeon and R. Metzler, \emph{ibid.} \textbf{81}, 021103
(2010).

\bibitem{jochen} J. Kursawe, J. H. P. Schulz, and R. Metzler, E-print
arXiv:1307.6131; J-.H. Jeon, N. Leijnse, L. Oddershede, and R. Metzler,
New J. Phys. \textbf{15}, 045011 (2013); J.-H. Jeon and R. Metzler, Phys.
Rev. E \textbf{85}, 021147 (2012).

\bibitem{samko} S. G. Samko, A. A. Kilbas, and O. O. I. Marichev,
\textit{Fractional integrals and derivatives} (Gordon and Breach, New York, NY,
1993).

\bibitem{yu1994} Y.-K. Yu, N.-N. Pang, and T. Halpin-Healy, Phys. Rev.
E \textbf{50}, 5111 (1994).

\bibitem{pang1997} N.-N. Pang, Phys. Rev. E \textbf{56}, 6676 (1997).

\bibitem{pang2010} N.-N. Pang and W.-J. Tzeng, Phys. Rev. E \textbf{82},
031605
(2010).

\bibitem{krug1997} J. Krug, H. Kallabis, S. Majumdar, S. Cornell, A.
Bray, and C. Sire, Phys. Rev. E \textbf{56}, 2702 (1997).

\bibitem{majumdar} S. N. Majumdar and A. J. Bray, Phys. Rev. Lett.
\textbf{86}, 3700 (2001).

\bibitem{kilbas} A. A. Kilbas, H. M. Srivastava, and J. J. Trujillo,
\textit{Theory and Applications of Fractional Differential Equations},
vol. 204 (Elsevier, 2006).

\bibitem{yang} Q. Yang, \textit{Novel analytical and numerical methods
for solving fractional dynamical systems} (Ph.D. Thesis, Queesland University
of Technology, Australia,
\href{http://eprints.qut.edu.au/35750}{http://eprints.qut.edu.au/35750}, 2010).

\bibitem{saichev1997} A. I. Saichev and G. M. Zaslavsky, Chaos
\textbf{7}, 753 (1997).

\bibitem{ilic2005} M. Ilic, F. Liu, I. Turner, and V. Anh, Fractl. Calc.
and Appl. Anal. \textbf{8}, 323 (2005); \emph{ibid.} \textbf{9}, 333 (2006).

\bibitem{zoia2007} A. Zoia, A. Rosso, and M. Kardar, Phys. Rev. E
\textbf{76}, 021116 (2007).

\bibitem{peitgen} H.-O. Peitgen, D. Saupe, M. F. Barnsley, Y. Fisher, and
M. McGuire, \textit{The science of fractal images} (Springer, New York, NY, 1988).

\bibitem{ausloos1985} M. Ausloos and D. Berman, Proc. Roy. Soc.
(London) A \textbf{400}, 331 (1985).

\bibitem{hamzehpour2006} H. Hamzehpour and M. Sahimi, Phys. Rev. E
\textbf{73}, 056121 (2006).

\bibitem{makse1996} H. A. Makse, S. Havlin, M. Schwartz, and H. E.
Stanley, Phys. Rev. E \textbf{53}, 5445 (1996).

\bibitem{taqqu} M. S. Taqqu and G. Samorodnisky, \emph{Stable non-Gaussian
random processes} (Chapman and Hall, New-York, NY, 1994).

\bibitem{jae} J.-H. Jeon and R. Metzler, J. Phys. A \textbf{43}, 252001 (2010).

\end{thebibliography}
\end{document}